\documentclass[a4paper,11pt]{article}
\usepackage{pos}
\usepackage[]{caption}
\usepackage{subcaption}
\usepackage{amsmath,amssymb}
\usepackage{wrapfig}
\usepackage{bm}
\usepackage{lineno}
\title{Characterization of the PeV astrophysical neutrino energy spectrum with IceCube using down-going tracks}
\ShortTitle{PeV down-going neutrino analysis}

\author{The IceCube Collaboration \\{\normalsize \normalfont(a complete list of authors can be found at the end of the proceedings)}}

\emailAdd{yanglyu@lbl.gov}

\abstract{ 
The IceCube Neutrino Observatory has observed a diffuse flux of astrophysical neutrinos with energies from TeV to a few PeV. Recent IceCube analyses have limited sensitivity to PeV neutrinos because upward-going neutrino fluxes are attenuated by the Earth while the Extremely High Energy (EHE) result targets cosmogenic neutrinos only above 10 PeV. In this work, we present a new event selection that fills the gap between 1 PeV and 10 PeV. This sample is obtained by selecting high-energy down-going through-going tracks from 8 years of data. To reduce the atmospheric muon backgrounds and achieve a high signal-to-background ratio, we combine two techniques. The first technique selects events with high stochasticity because single muons created by neutrinos lose energy more stochastically than atmospheric muon bundles whose energy losses are smoothened due to large muon multiplicities. The second technique uses the IceTop surface array as a veto of atmospheric background events. To characterize the astrophysical neutrino flux and test the existence of a cut-off in the neutrino energy spectrum at a few PeV, a global fit will be performed by combining this sample with results from the 7-year High Energy Starting Events (HESE) analysis. 

\vspace{4mm}
{\bfseries Corresponding author:}
Yang Lyu$^{1,2*}$\\
{$^{1}$ \itshape University of California, Berkeley\\
$^{2}$ Lawrence Berkeley National Laboratory}\\
$^*$ Presenter

\FullConference{37$^{\rm{th}}$ International Cosmic Ray Conference (ICRC 2021)\\
		July 12th -- 23rd, 2021\\
		Online -- Berlin, Germany}
}

\begin{document}
\maketitle 

% %%%%%%%%%%%%%%%%%%%%%%%%%%%%%%% BEGIN HERE %%%%%%%%%%%%%%%%%%%%%%%%%%%%%%% 

\section{Introduction}\label{sec:introduction}

The origins and acceleration mechanisms of Ultra-High Energy Cosmic Rays (UHECRs) remain a mystery. Astrophysical neutrinos created in or near the acceleration sites of cosmic rays could give us insights into these questions \cite{Gaisser:1995-particle-astrophysics-with-neutrino}. 

IceCube is a cubic-kilometer neutrino detector installed in the ice at the geographic South Pole \cite{Aartsen:2016nxy} with an accompanying air shower array, IceTop, on the surface of the ice \cite{Abbasi:2013-IceTop-surface-array}. The first observation of the diffuse astrophysical neutrino flux at IceCube was announced in 2013, where high-energy events with interaction vertices contained inside the detector were studied \cite{Aartsen:2013-evidence-for-extraterrestrial-neutrinos}. A later analysis achieved a larger effective volume by analyzing muon tracks created by $\nu_\mu$ events from the northern hemisphere, where the interaction vertices could be outside the detector \cite{Aartsen:2015-first-northern-track}. These two event selections are most sensitive to neutrinos with energies below 1 PeV because the Earth is opaque to neutrinos beyond this energy. A third analysis selects for extremely-high-energy tracks from the southern sky by applying strong cuts to remove atmospheric muon backgrounds, and it is sensitive to neutrinos above 10 PeV \cite{Ishihara:2016-EHE}. 

To fill the gap between 1 PeV and 10 PeV, we present a new event selection that targets high-energy down-going tracks where atmospheric muon backgrounds are removed using the stochasticity information and IceTop as a veto. The muon backgrounds in the signal region are estimated with a stochasticity template fitting technique. The signal events we obtain will be at high energy and have angular resolutions of 0.2$^\circ$. To measure the astrophysical neutrino flux and energy spectrum, we will combine our sample with the 7.5-year HESE sample \cite{Schneider:2019-HESE-7yr} to perform a combined fit. Our sample will also be used for point source searches in future studies.

Characterizing the highest-energy astrophysical neutrino flux could provide us insights into the PeV spectral cutoff. The deficiency of observed events above 2 PeV could indicate an exponential cutoff in the neutrino energy spectrum. Many neutrino source models predict such a cutoff due to mechanisms such as synchrotron pion cooling \cite{Petropoulou:2014-cutoff-pion-cooling}, limited source energies \cite{Becker:2009-cutoff-limited-source-energy}, etc. By performing a combined fit, we may better constrain the spectral cutoff energy and the production models of neutrinos and cosmic rays.

\section{Event Selection}\label{sec:event_selection}

In this analysis, we use 8 years of burn sample (10\% of the full dataset) from 2012 to 2019. To ensure the neutrinos in our final sample are both at high energy and track-like, we first apply initial selections including reconstructed muon energy > 300 TeV (high-energy), zenith < 90$^{\circ}$ (down-going), and track lengths > 600 m (track-like).

\begin{figure}[htbp]
\centering
\begin{subfigure}{.45\textwidth}
  \centering
  \includegraphics[width=0.95\linewidth]{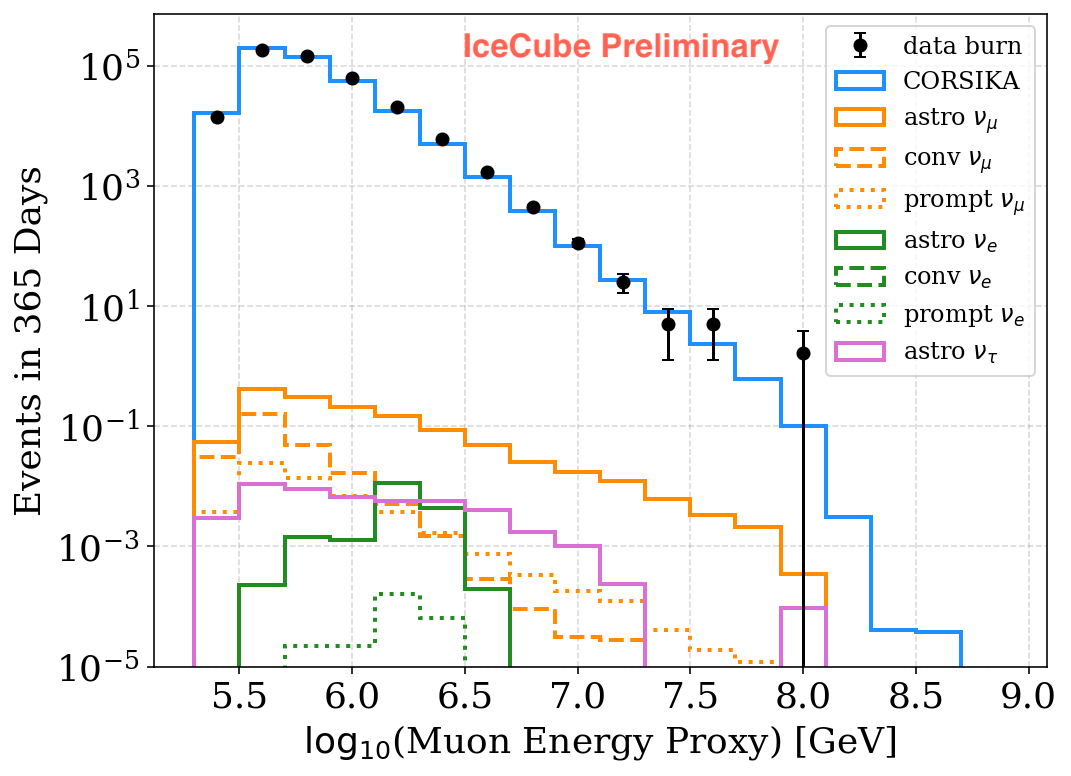}
%   \caption{data}
%   \label{fig:sub1}
\end{subfigure}%
\begin{subfigure}{.45\textwidth}
  \centering
  \includegraphics[width=0.95\linewidth]{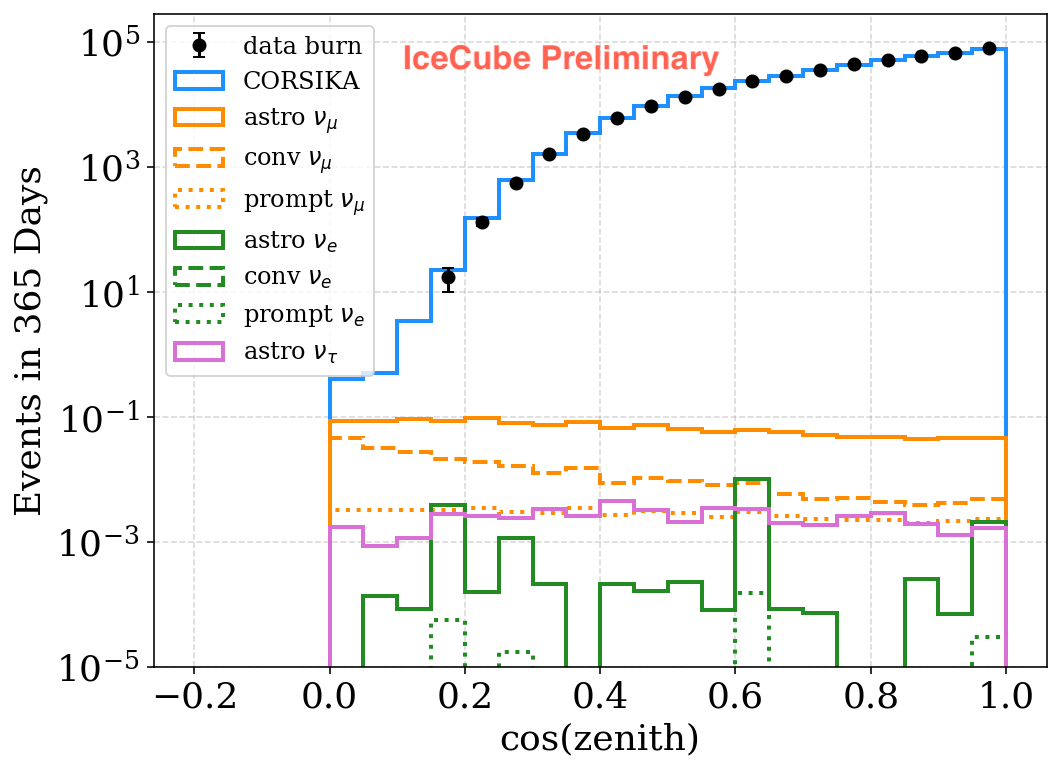}
%   \caption{model}
%   \label{fig:sub2}
\end{subfigure}
\caption{The muon energy and zenith distributions after initial selections. Conventional atmospheric neutrinos from kaon and pion decays are weighted according to Honda \textit{el al.} \cite{Honda:2006}, and prompt components from charm meson decays are weighted to the BERSS calculation \cite{Bhattacharya:2015-BERSS}. An enhancement of the conventional $\nu_e$ flux from $K_s$ at high energy  \cite{Gaisser-2015-newflux} could potentially be added in a future model.}
\label{fig:after-presel}
\end{figure}

Fig.~\ref{fig:after-presel} shows the distributions of muon energy and zenith for data and simulations after the initial selections, normalized to one year. The atmospheric muon background is simulated with CORSIKA and is weighted to the Gaisser-Hillas model (H3a) \cite{Gaisser:2012-Gaisser-Hillas-H3a}, and neutrinos are simulated with neutrino-generator \cite{ANIS:2005-nugen}. In addition to $\nu_\mu$, a small contribution from $\nu_e$ is observed because muons are produced in hadronic showers after the deep inelastic scattering. A contribution from $\nu_\tau$ comes from $\tau$ leptons decaying to muons. The all-flavor $\nu + \bar{\nu}$ astrophysical neutrino fluxes are assumed to be $\frac{d\Phi_{6\nu}}{dE} = \Phi_{\mathrm{astro}} \cdot \Big(\frac{\mathrm{E}}{100\mathrm{TeV}}\Big)^{-\gamma_{\mathrm{astro}}} \cdot 10^{-18}$ GeV$^{-1}$s$^{-1}$sr$^{-1}$cm$^{-2}$, where the six-neutrino flux normalization $\Phi_{\mathrm{astro}}=6.7$ and the spectral index $\gamma_{\mathrm{astro}}=2.5$, as in the previous combined-fit results \cite{Aartsen:2015-combined-fit}. The final analysis will be performed with a binned Poisson likelihood fit with 20 energy bins and 10 zenith bins. 

\subsection{Stochasticity Measurement}\label{sec:stochasticity}

\setlength{\intextsep}{0pt}%
\setlength{\columnsep}{10pt}%
\begin{wrapfigure}{R}{0.4\textwidth}
  \centering
    \includegraphics[width=0.4\textwidth]{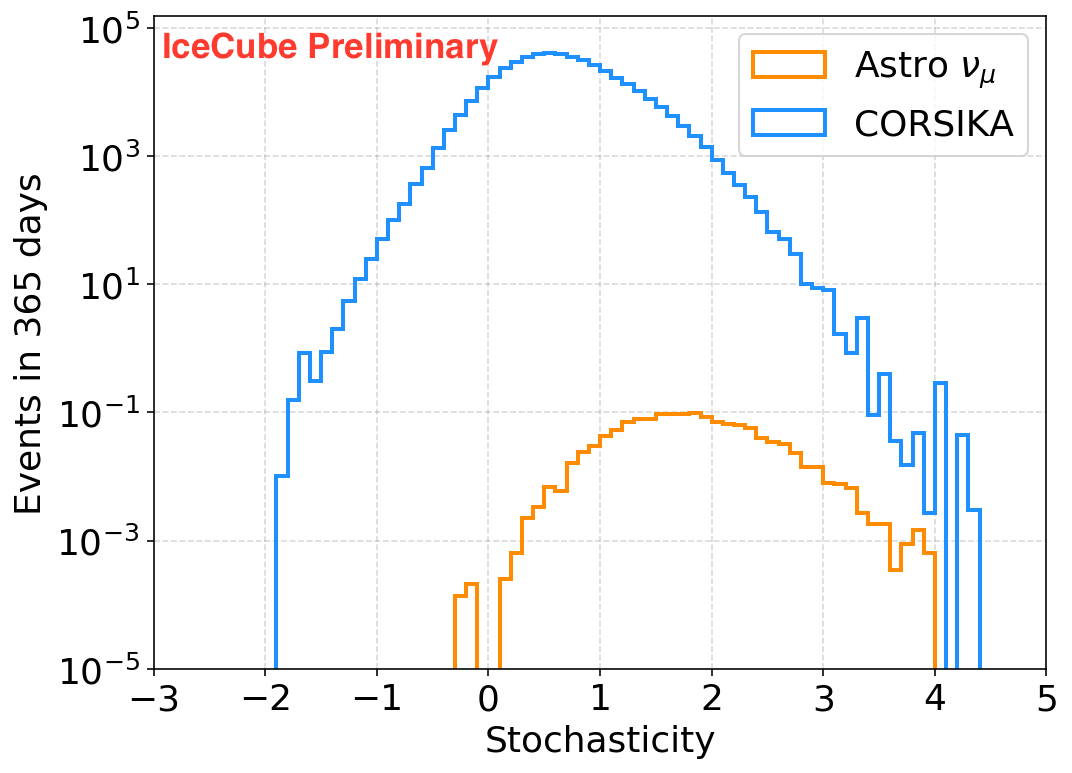}
  \caption{Stochasticity distributions of signal ($\nu_\mu$) vs. backgrounds (CORSIKA). In practice, the stochasticity proxy we use is $\log_{10}\mathrm{MSE}$, and its value could be negative. The median stochasticity for CORSIKA is 0.6 and for $\nu_\mu$ is 2.0.}
  \label{fig:stochasticity}
\end{wrapfigure}

As we have seen in the event distributions, the atmospheric muon background is several orders of magnitude higher than the neutrino signals. The stochasticity of the muon track is a powerful tool to reduce this background. Single muons created by $\nu_\mu$ lose energy stochastically at high energies. On the other hand, atmospheric muon bundles have smoother observed energy losses because of the large muon multiplicity in each bundle. 

To measure the stochasticity of the muon, we divide the muon track into 120-meter-wide segments and calculate the energy losses for each segment \cite{Abbasi:2012-truncated-energy}. We then fit a line to the energy losses and use the weighted mean squared error (MSE) as the proxy for stochasticity. The weight for each energy loss $\mathrm{E_i}$ is assumed to be $\sqrt{\mathrm{E_i}}$, a choice optimized with Monte Carlo simulations, to maximize the separation power between signal and background. Fig.~\ref{fig:stochasticity} shows the stochasticity distribution of CORSIKA and $\nu_\mu$ events, where astrophysical $\nu_\mu$ peaks at higher stochasticity.

\subsection{IceTop Veto}\label{sec:IceTop_veto}

\begin{wrapfigure}{R}{0.35\textwidth}
  \centering
    \includegraphics[width=0.35\textwidth]{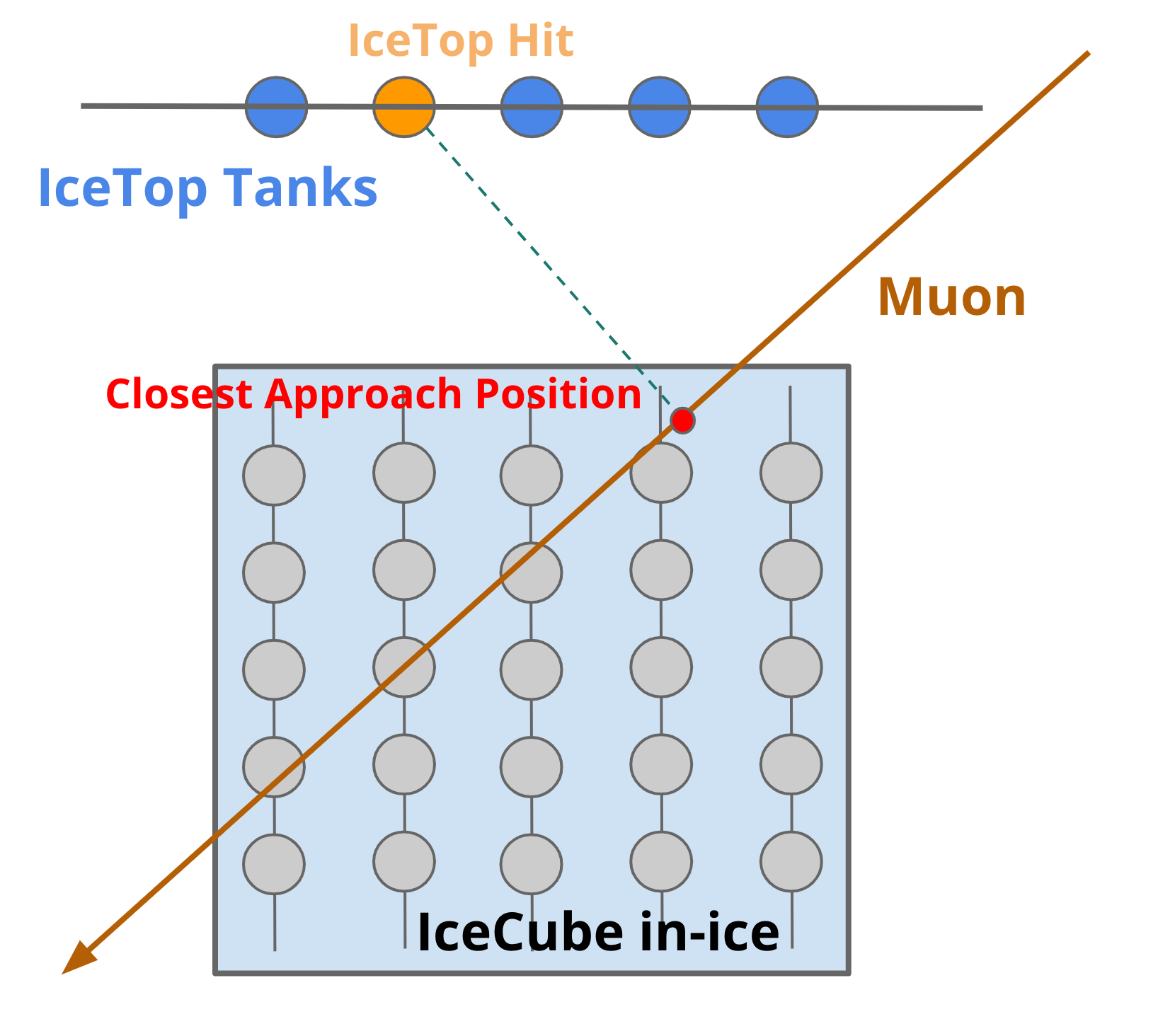}
  \caption{IceTop veto illustrated.}
  \label{fig:veto-illustration}
\end{wrapfigure}

Another handle to reduce the atmospheric muon backgrounds is to use the IceTop array as a veto. Atmospheric muons are part of a cosmic ray air shower that may deposit charges (recorded as hits) in IceTop tanks. If an IceTop tank is hit, we calculate the time when the muon arrives at the closest position relative to that tank. If the difference between the hit time and the closest-approach time is within the time window [-700 ns, 1700 ns], then this hit is correlated with the event. We veto the event if there are $\geq$ 2 correlated IceTop hits in total. The veto principle is illustrated in Fig.~\ref{fig:veto-illustration}. The width of the time window is optimized by examining the calculated time differences for all IceTop hits in the burn sample. In addition, using a background window at [-4500 ns, -2100 ns], the random-hit rate is found to be ~0.5 per event, so a veto threshold of 2 hits is justified.

\setlength{\intextsep}{5pt}%
\setlength{\columnsep}{10pt}%

This veto approach may fail for some down-going events. If an atmospheric muon has a large zenith angle or has low energy, IceTop may not observe enough hits from the accompanying air shower to veto the event. Therefore, we perform a data-driven modeling for the effectiveness of IceTop and only apply veto in the region where IceTop is effective enough. We use the full dataset in the background region (stochasticity < 0.8) for this modeling to reduce bias from neutrino signals. The effectiveness is parametrized as the ``inefficiency,'' which is a function of muon energy and distance from the muon track to the center of IceTop. 

\begin{figure}[htbp]
\centering
\begin{subfigure}{.45\textwidth}
  \centering
  \includegraphics[width=0.95\linewidth]{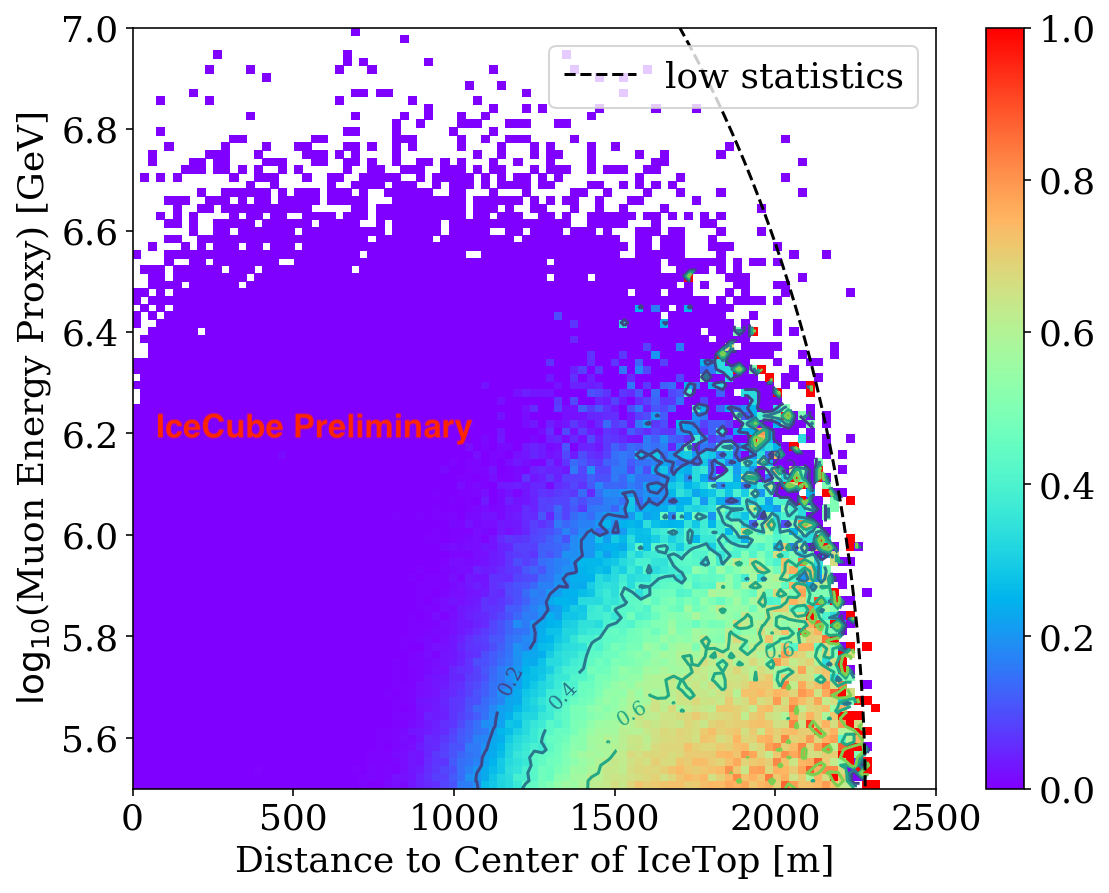}
  \caption{IceTop inefficiency calculated from data.}
  \label{fig:IT-ineff-1}
\end{subfigure}%
\begin{subfigure}{.45\textwidth}
  \centering
  \includegraphics[width=0.95\linewidth]{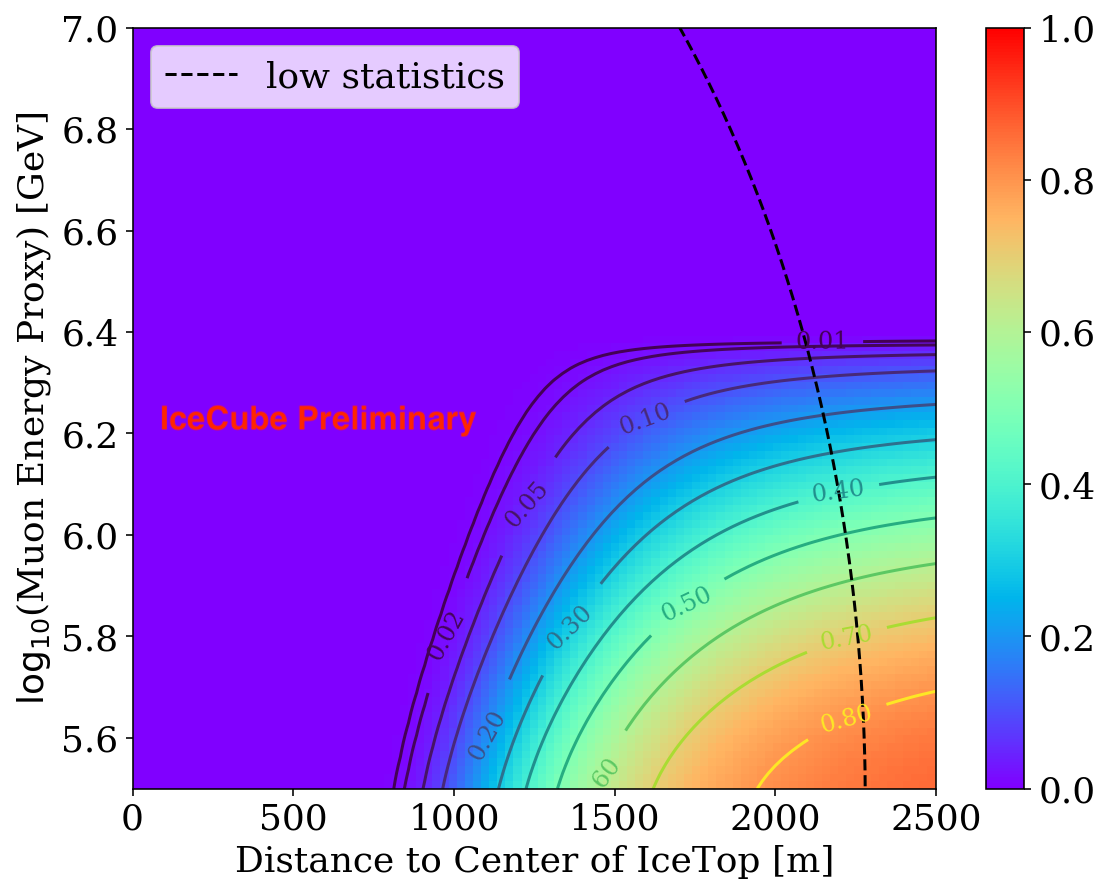}
  \caption{IceTop inefficiency model.}
  \label{fig:iT-ineff-2}
\end{subfigure}
\caption{IceTop inefficiency calculated from data in the background region (stochasticity < 0.8) and its modeling. Colored curves corresponds to inefficiency contours. The region to the right of the black dotted line has low data statistics. In practice, inefficiency beyond this dashed line is manually set to 1.}
\label{fig:IT-ineff}
\end{figure}

Fig.~\ref{fig:IT-ineff-1} shows the inefficiency calculated from data. In each energy-distance bin, inefficiency is defined as the fraction of events that are not vetoed by IceTop. Therefore, a smaller inefficiency indicates a better bundle rejection power. Fig.~\ref{fig:iT-ineff-2} shows the inefficiency model obtained by fitting a function of three variables to data on the left. Values in low-statistics regions are extrapolated. Even at large distances (corresponding to large zenith angles on average), IceTop could still veto over 99\% of atmospheric muons if the muon energy is high. In addition to atmospheric muons, atmospheric neutrino fluxes are suppressed further due to accompanying air showers.

\subsection{Background Estimation}\label{sec:bg_estimation}

The signal region is defined by a series of cuts: 1) initial selections, 2) stochasticity > $c_1$, 3) IceTop inefficiency < $c_2$, and 4) pass IceTop veto. After these cuts, the signal region will still contain some background events, mostly from atmospheric muons. Part of the backgrounds is due to the imperfect IceTop veto. In addition, proton primaries may create events where a single muon dominates, leading to an indistinguishable background based on stochasticity. We optimize $c_1$ and $c_2$ such that we maximize the number of events in the signal region while requiring signal-to-background ratio (S/B) > 2. We require a high S/B to increase the sensitivity to a PeV cutoff and to increase the sample purity for the future point source analysis.

\setlength{\intextsep}{0pt}%
\setlength{\columnsep}{10pt}%
\begin{wrapfigure}{R}{0.4\textwidth}
  \centering
    \includegraphics[width=0.4\textwidth]{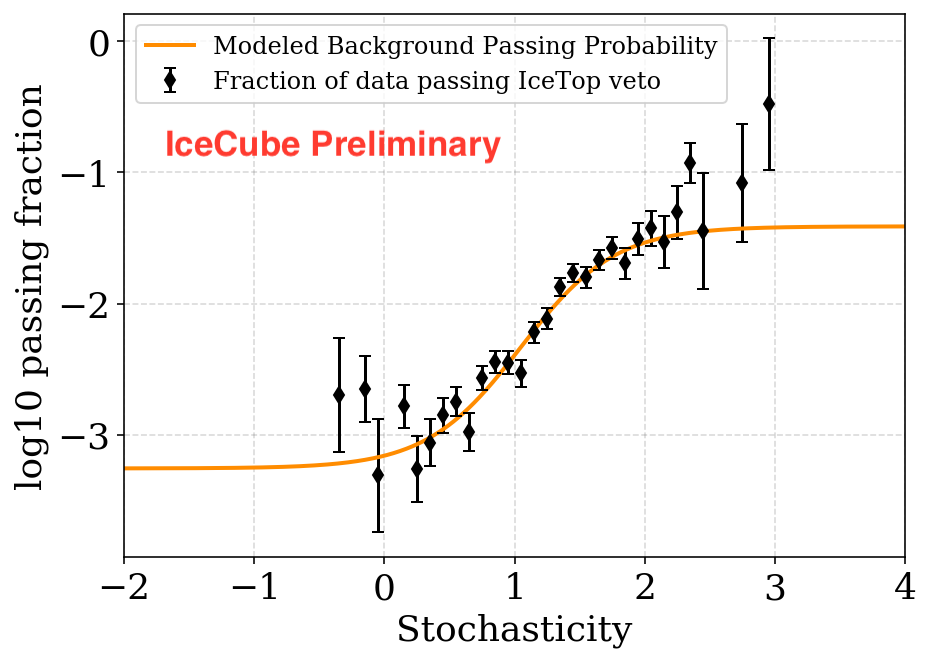}
  \caption{Correlation between IceTop hits and stochasticity. Black: passing fraction of the burn sample. Orange: modeled passing probability of atmospheric muons using CORSIKA. }
  \label{fig:passing-fraction}
\end{wrapfigure}

However, estimating these remaining backgrounds is challenging due to a correlation between stochasticity and IceTop hits. In Fig.~\ref{fig:passing-fraction}, each black marker shows the fraction of data passing the IceTop veto as a function of stochasticity inside the region with inefficiency < 0.01. The observed positive correlation between the passing fraction and stochasticity is studied using CORSIKA. For bins at higher stochasticity, the fraction of lighter elements, such as protons, is higher. On the other hand, protons on average have smaller air shower energies comparing to heavier elements such as iron. Therefore, events at higher stochasticity are accompanied by fewer IceTop hits on average and are easier to pass the veto. 

To estimate the remaining backgrounds, we model the passing probability of atmospheric muons as a function of stochasticity by performing a template fitting using CORSIKA. We first split the CORSIKA events into ``singles'' and ``bundles'' according to the singleness parameter: $\epsilon = E_{\mathrm{leading\ muon}}/E_{\mathrm{bundle}}$ and refer their stochasticity distributions as templates. Then we fit the two templates to data to obtain the correct normalization. Finally, we parametrize the passing probability of atmospheric muons using the two templates and fit the model to data in the background region. The best-fit model is the orange curve in Fig.~\ref{fig:passing-fraction}. The discrepancy between data and model at high stochasticity is expected due to the presence of neutrino signal events. The background in the signal region is then calculated on an event-by-event basis using the best-fit model.

\section{Burn Sample Events Passing the Selection}

\setlength{\intextsep}{0pt}%
\setlength{\columnsep}{10pt}%
\begin{wrapfigure}{R}{0.4\textwidth}
  \centering
    \includegraphics[width=0.4\textwidth]{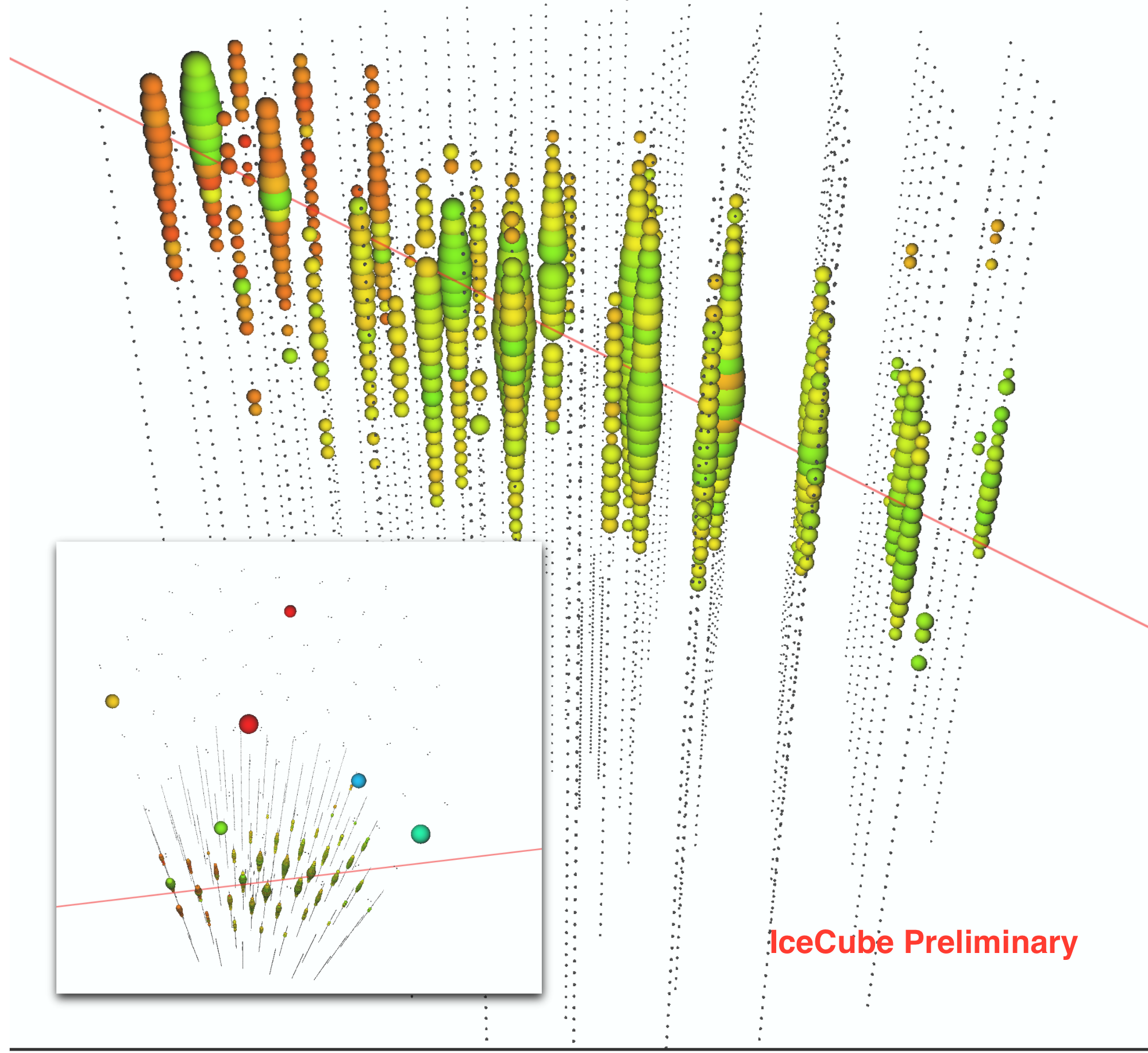}
  \caption{Visualization of one burn sample event in the signal region.}
  \label{fig:visualization}
\end{wrapfigure}

The best cuts that maximize the number of events in the signal region while maintaining the S/B > 2 are 1) stochasticity > 2.5 and 2) IceTop inefficiency < 0.011. Two burn sample events pass the cuts with estimated S/B = 2.17. The event-wise backgrounds are 0.58 and 0.11, respectively. After unblinding, we expect to obtain around 20 events with a high signal-to-background ratio.

Fig.~\ref{fig:visualization} shows a visualization of one of the signal events. The size of each circle represents the amount of charge recorded for each DOM; red and green colors indicate early and late photon arrival times. The red track shows a muon traveling from left to right; the muon energy at the center of the detector is estimated to be $3.2\substack{+2.1 \\ -1.3}$ PeV with a zenith angle of 69$^\circ$. The small panel on the lower left shows the event viewed from above with IceTop displayed. There are several accompanying IceTop hits, but all of them except one are outside the veto time window. The other event (not shown) is at 40$^\circ$ with muon energy of $1.2\substack{+0.7\\-0.5}$ PeV, and one out of five IceTop hits is correlated.

\section{Flux Characterization}\label{sec:flux}

After unblinding, we will combine our sample with the 7.5-year HESE sample to perform a combined fit. The binned Poisson likelihood function is constructed as
\begin{equation}
    \ln L(\pmb{\theta}, \pmb{\xi}) =\sum_{i} \ln\Big(\frac{e^{-\mu_i}\mu_i^{k_i}}{k_i!}\Big) + \sum_{j}^{}\Big(\frac{\xi_j - \xi_j^*}{\sigma[\xi_j]}\Big)^2
\end{equation}
The likelihood is a function of physics parameters $\pmb{\theta}$ (such as $\Phi_{\mathrm{astro}}$, $\gamma_{\mathrm{astro}}$, and the spectral cutoff energy $E_{\mathrm{cutoff}}$) and nuisance parameters $\pmb{\xi}$ associated with parametrized systematic uncertainties. The systematic uncertainties are associated with the detector response, the ice properties, the flux models, and the background estimation procedure. Confidence intervals for best-fit values will be calculated assuming Wilk's theorem \cite{Wilks:1938}. Parameters that are shared both in HESE and this sample take the same values for both samples. The first part of the likelihood is the Poisson term summed over analysis bins $i$ in energy and zenith; $\mu_i(\pmb{\theta}, \pmb{\xi})$ denotes the expectation from Monte Carlo simulations and background estimations, and $k_i$ denotes the observed data. The HESE sample is binned in energy, zenith, and morphology; this analysis is primarily binned in energy and zenith, but a third dimension of event morphology is concatenated (assigning all events as tracks) to perform a joint fit. The second part of the likelihood penalizes unphysical results. It sums over nuisance parameters $\xi_j$, each with a prior $\xi_j^*$ and an uncertainty
$\sigma[\xi_j]$.

To study the preference of the generic spectral cutoff flux model over the single power law (SPL) model, we will perform a likelihood ratio test by scanning over $E_{\mathrm{cutoff}}$ and calculate the test statistic
\begin{equation}
    TS(E_{\mathrm{cutoff}}) = -2\ln \frac{L_{\mathrm{cutoff}}(E_{\mathrm{cutoff}},\hat{\hat{\pmb{\eta}}})}{L_{\mathrm{SPL}}(\hat{\pmb{\eta}})}
\end{equation} 
where parameters $\hat{\pmb{\eta}}$ maximize $L_{\mathrm{SPL}}$ and $\hat{\hat{\pmb{\eta}}}$ maximize $L_{\mathrm{cutoff}}$ for each fixed $E_{\mathrm{cutoff}}$. The spectral cutoff model differs from the SPL model by an additional exponential term
\begin{equation}
    \frac{d\Phi_{\mathrm{cutoff}}}{dE} = \Phi_{\mathrm{astro}} \cdot \Big(\frac{\mathrm{E}}{100\mathrm{TeV}}\Big)^{-\gamma_{\mathrm{astro}}}\cdot \mathrm{e}^{-\frac{E}{E_{\mathrm{cutoff}}}} \cdot 10^{-18}\cdot \mathrm{GeV}^{-1}\mathrm{s}^{-1}\mathrm{sr}^{-1}\mathrm{cm}^{-2}
\end{equation}
The p-value for rejecting the null hypothesis (SPL) and the favored region for the cutoff energy will be calculated.

\section{Systematic Uncertainties}\label{sec:syst}

\subsection{Background Estimation Uncertainties}\label{sec:syst_bg}

\begin{wrapfigure}{R}{0.4\textwidth}
  \centering
    \includegraphics[width=0.4\textwidth]{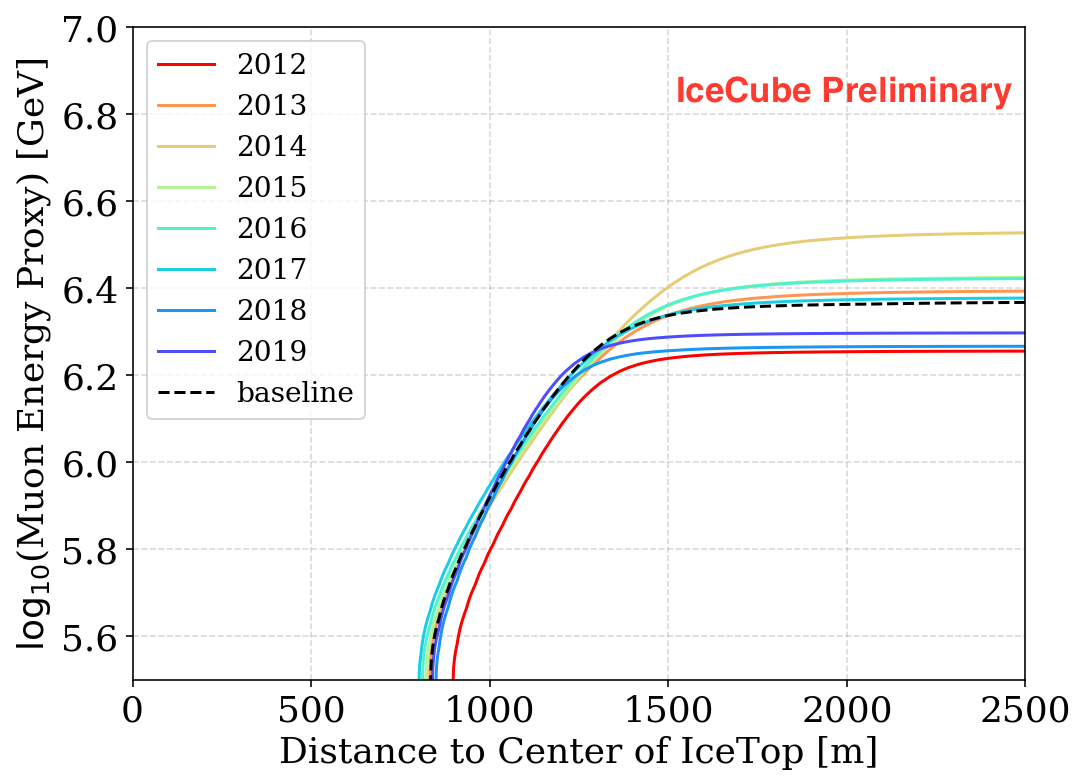}
  \caption{IceTop inefficiency contours (at 0.01) modeled using different years of data. The baseline corresponds to the modeling with the full dataset. The large fluctuations of the contour lines at high energy comes from 1) low statistics of data 2) an artifact due to the functional form chosen for the modeling process, leading to an overestimation of the systematic uncertainties at high energy.}
  \label{fig:snow-effect}
\end{wrapfigure}

There are several systematic uncertainties associated with the background estimation modeling. First, as snow slowly accumulates on the surface of the ice, the sensitivity of IceTop decreases, and the modeled inefficiency contours shift as shown in Fig.~\ref{fig:snow-effect}. Therefore, after applying inefficiency cuts on the CORSIKA sample, the shapes of stochasticity templates and estimated backgrounds will be affected. To quantify this snow effect, we find two limiting inefficiency contours from the baseline model to capture the variations for different years of models. We then perform the CORSIKA template fitting and re-estimate the backgrounds for the two burn sample events in the signal region. The uncertainty due to the snow effect is found to be around 5\%. 

\setlength{\intextsep}{5pt}%
\setlength{\columnsep}{10pt}%

The second uncertainty comes from the stochasticity cut (stochasticity < 0.8 as baseline) applied when modeling the IceTop inefficiency. Because stochasticity is correlated with IceTop hits, different stochasticity thresholds will lead to different IceTop inefficiency contours. To quantify this uncertainty, we apply different cut values from 0.4 to 1.2 and perform the template fitting for each scenario and estimate the backgrounds. The uncertainty due to the stochasticity cut is around 2\%. 

The third uncertainty comes from the singleness threshold that we use to separate CORSIKA events into ``single'' and ``bundle'' templates. Similarly, we vary the threshold (singleness = 0.8 as baseline) from 0.7 to 0.9 and re-estimate the backgrounds. The uncertainty due to the singleness parameter is around 20\%. This is the dominant uncertainty and is parametrized as a nuisance parameter to scale the backgrounds for each signal event.

Lastly, we estimate the backgrounds using different CORSIKA fluxes, H3a and H4a, where H4a \cite{Gaisser:2012-Gaisser-Hillas-H3a} assumes that cosmic rays of extra-galactic population are all protons. The associated background estimation uncertainty is less than 1\%. 

\subsection{Detector Systematics}\label{sec:syst_det}

There are three detector and ice systematic uncertainties that would affect the analysis, and they are parametrized as continuously-varying nuisance parameters. $\epsilon_{\mathrm{DOM}}$ parametrizes the overall optical efficiency of all DOMs. An increase in $\epsilon_{\mathrm{DOM}}$ shifts the peak of the energy distribution to higher values. The effect on the zenith distribution is to scale up the overall normalization. $\epsilon_{\mathrm{absorption}}$ parametrizes the bulk-ice absorption and scattering effects. However, only the absorption coefficient of the bulk ice affects the energy or zenith distributions. Lastly, $\epsilon_{\mathrm{hole}}$ describe the local angular acceptance of DOMs. It is parametrized to only scale the overall normalization of the distributions due to the low statistics in the simulations. The parametrizations are performed using Monte Carlo simulations with different detector or ice property settings.

\subsection{Neutrino Flux Uncertainties}\label{sec:syst_flux}

We use $\Phi_{\mathrm{conv}}$ and $\Phi_{\mathrm{prompt}}$ to scale normalizations of the conventional and prompt atmospheric neutrino fluxes. Due to the low contribution from atmospheric neutrinos, we ignore uncertainties such as the kaon-to-pion ratio, the hardening or softening in the neutrino energy spectrum due to cosmic ray flux uncertainties, and the hadronic production uncertainties.

\section{Conclusions}\label{sec:conclusions}

We have developed a new analysis to fill the neutrino energy gap between 1 PeV and 10 PeV. We use both stochasticity cuts and IceTop veto to reduce atmospheric muon backgrounds. The backgrounds in the signal region are estimated with a stochasticity template fitting method. We observe 2 burn sample events in the signal region and expect to obtain ~20 high-energy events after unblinding. The final sample will be dominated by astrophysical neutrinos with an estimated signal-to-background ratio above 2. This sample will be combined with the HESE sample to perform a combined fit, and the existence of a PeV cutoff will be tested. Because of the high purity and excellent angular resolution, this sample will also be used for point source searches in future studies.

% %%%%%%%%%%%%%%%%%%%%%%%%%%%%%%% END HERE %%%%%%%%%%%%%%%%%%%%%%%%%%%%%%% 

% Set up the bibliography using BibTeX.
% Get references from inspirehep.net or NASA/ADS and put them in references.bib.
\bibliographystyle{ICRC}
\bibliography{references}

% Full authors list (ONLY FOR COLLABORATIONS)
\clearpage
\section*{Full Author List: IceCube Collaboration}

% \noindent \textbf{Note comment afterwards:} Collaborations have the possibility to provide an authors list in xml format which will be used while generating the DOI entries making the full authors list searchable in databases like Inspire HEP. For instructions please go to icrc2021.desy.de/proceedings or contact us under icrc2021proc@desy.de.\\

% \scriptsize
% \noindent
% first.author$^1$, 
% second.author$^2$, 
% third.author$^3$ % .... more names
% and 
% last.author$^{n}$ \\

% \noindent
% $^1$first.affiliation.
% $^2$second.affiliation. % .... more affiliation
% $^{m}$last.affiliation.

\scriptsize
\noindent
R. Abbasi$^{17}$,
M. Ackermann$^{59}$,
J. Adams$^{18}$,
J. A. Aguilar$^{12}$,
M. Ahlers$^{22}$,
M. Ahrens$^{50}$,
C. Alispach$^{28}$,
A. A. Alves Jr.$^{31}$,
N. M. Amin$^{42}$,
R. An$^{14}$
% ...
\\

\noindent
% ...
$^{12}$ Universit{\'e} Libre de Bruxelles, Science Faculty CP230, B-1050 Brussels, Belgium \\
% ...
$^{14}$ Department of Physics and Laboratory for Particle Physics and Cosmology, Harvard University, Cambridge, MA 02138, USA \\
% ...
$^{17}$ Department of Physics, Loyola University Chicago, Chicago, IL 60660, USA \\
$^{18}$ Dept. of Physics and Astronomy, University of Canterbury, Private Bag 4800, Christchurch, New Zealand \\
% ...
$^{22}$ Niels Bohr Institute, University of Copenhagen, DK-2100 Copenhagen, Denmark \\
% ...
$^{28}$ D{\'e}partement de physique nucl{\'e}aire et corpusculaire, Universit{\'e} de Gen{\`e}ve, CH-1211 Gen{\`e}ve, Switzerland \\
% ...
$^{31}$ Karlsruhe Institute of Technology, Institute for Astroparticle Physics, D-76021 Karlsruhe, Germany  \\
% ...
$^{42}$ Bartol Research Institute and Dept. of Physics and Astronomy, University of Delaware, Newark, DE 19716, USA \\
% ...
$^{50}$ Oskar Klein Centre and Dept. of Physics, Stockholm University, SE-10691 Stockholm, Sweden \\
% ...
$^{59}$ DESY, D-15738 Zeuthen, Germany \\
% ...

\subsection*{Acknowledgements}

\noindent
USA {\textendash} U.S. National Science Foundation-Office of Polar Programs,
U.S. National Science Foundation-Physics Division,
U.S. National Science Foundation-EPSCoR,
Wisconsin Alumni Research Foundation,
Center for High Throughput Computing (CHTC) at the University of Wisconsin{\textendash}Madison,
Open Science Grid (OSG),
Extreme Science and Engineering Discovery Environment (XSEDE),
Frontera computing project at the Texas Advanced Computing Center,
U.S. Department of Energy-National Energy Research Scientific Computing Center,
Particle astrophysics research computing center at the University of Maryland,
Institute for Cyber-Enabled Research at Michigan State University,
and Astroparticle physics computational facility at Marquette University;
Belgium {\textendash} Funds for Scientific Research (FRS-FNRS and FWO),
FWO Odysseus and Big Science programmes,
and Belgian Federal Science Policy Office (Belspo);
Germany {\textendash} Bundesministerium f{\"u}r Bildung und Forschung (BMBF),
Deutsche Forschungsgemeinschaft (DFG),
Helmholtz Alliance for Astroparticle Physics (HAP),
Initiative and Networking Fund of the Helmholtz Association,
Deutsches Elektronen Synchrotron (DESY),
and High Performance Computing cluster of the RWTH Aachen;
Sweden {\textendash} Swedish Research Council,
Swedish Polar Research Secretariat,
Swedish National Infrastructure for Computing (SNIC),
and Knut and Alice Wallenberg Foundation;
Australia {\textendash} Australian Research Council;
Canada {\textendash} Natural Sciences and Engineering Research Council of Canada,
Calcul Qu{\'e}bec, Compute Ontario, Canada Foundation for Innovation, WestGrid, and Compute Canada;
Denmark {\textendash} Villum Fonden and Carlsberg Foundation;
New Zealand {\textendash} Marsden Fund;
Japan {\textendash} Japan Society for Promotion of Science (JSPS)
and Institute for Global Prominent Research (IGPR) of Chiba University;
Korea {\textendash} National Research Foundation of Korea (NRF);
Switzerland {\textendash} Swiss National Science Foundation (SNSF);
United Kingdom {\textendash} Department of Physics, University of Oxford.

\end{document}